\documentclass[10pt,technote,twocolumn]{IEEEtran}
\usepackage{amsmath,amsfonts}
\usepackage{algorithmic}
\usepackage{algorithm}
\usepackage{array}
\usepackage[caption=false,font=normalsize,labelfont=sf,textfont=sf]{subfig}
\usepackage{textcomp}
\usepackage{stfloats}
\usepackage{url}
\usepackage{verbatim}
\usepackage{graphicx}
\usepackage{cite}
\usepackage{makecell}
\usepackage{xcolor}
\newcommand{\rev}[1]{#1}
\ifdefined\TVTmarked
  \newcommand{\trackchange}[1]{{\color{red}#1}}
\else
  \newcommand{\trackchange}[1]{{\color{black}#1}}
\fi
\begin{document}

\title{\rev{\trackchange{R2WM}: Recursive Reasoning Wireless Model for \trackchange{Long-Horizon mmWave} Beam Prediction}}

\author{Yunxiao~Tian, Li~Qiao,~\IEEEmembership{Member,~IEEE,} Ziwei~Wan,~\IEEEmembership{Member,~IEEE,} Andrey~Ivanov, and Zhen~Gao,~\IEEEmembership{Senior~Member,~IEEE}%
\thanks{\trackchange{\copyright~2026 IEEE. Personal use of this material is permitted.
Permission from IEEE must be obtained for all other uses, in any current or future media,
including reprinting/republishing this material for advertising or promotional purposes,
creating new collective works, for resale or redistribution to servers or lists,
or reuse of any copyrighted component of this work in other works.}}%
\thanks{\trackchange{The work was supported in part by Beijing Natural Science Foundation under Grant L242011, in part by the Natural Science Foundation of Shandong Province under Grant ZR2025QA30, in part by the National Natural Science Foundation of China (NSFC) under Grant 62471036, and in part by the Fundamental Research Funds for the Central Universities. \textit{(Corresponding~author: Li Qiao and Zhen Gao.)}}}%
\thanks{\trackchange{Y. Tian is with the School of Information and Electronics, Beijing Institute of Technology, Beijing, China (e-mail: yunxiao\_tian@bit.edu.cn).}}%
\thanks{L. Qiao is with the Department of Electrical and Computer Engineering, The University of Hong Kong, Pokfulam Road, Hong Kong (e-mail: qiaoli@hku.hk).}%
\thanks{Z. Wan is with the Yangtze Delta Region Academy, Beijing Institute of Technology (Jiaxing), Jiaxing, China (e-mail: ziweiwan@bit.edu.cn).}%
\thanks{A. Ivanov is with the Project Center for Next Generation Wireless and IoT, Skolkovo Institute of Science and Technology (Skoltech), Moscow, Russia (e-mail: AN.Ivanov@skoltech.ru).}%
\thanks{\trackchange{Zhen Gao is with Beijing Institute of Technology (BIT), Zhuhai 519088,
China, also with the State Key Laboratory of CNS/ATM, Beijing 100081,
China, also with the MIIT Key Laboratory of Complex-Field Intelligent
Sensing, Beijing 100081, China, also with the Advanced Technology Research
Institute, BIT, Jinan 250307, China, also with the Yangtze Delta Region
Academy, BIT, Jiaxing 314019, China, also with Zhengzhou Research Institute,
BIT, Zhengzhou 450000, China, and also with Chongqing Innovation
Center, BIT, Chongqing 401100, China (e-mail: gaozhen16@bit.edu.cn).}}%
}

\markboth{IEEE Transactions on Vehicular Technology}%
{Tian \MakeLowercase{\textit{et al.}}: R2WM: Recursive Reasoning Wireless Model for Long-Horizon mmWave Beam Prediction}

\IEEEpubid{\copyright~2026 IEEE}
\IEEEaftertitletext{\vspace{-12pt}}
\maketitle\label{rev:title}

\begin{abstract}
\leavevmode\label{rev:abstract}\rev{\trackchange{Long-horizon beam prediction aims to extend beam-training intervals while supporting beam alignment in mobile millimeter-wave systems. Although pretrained large models can improve prediction over compact recurrent baselines, their large backbones impose substantial storage and computing requirements. To address this limitation, we propose a recursive reasoning wireless model (R2WM). Given a fixed history of beam indices and angles of departure (AoDs), R2WM repeatedly updates coupled latent and prediction states using a shared causal-attention backbone before jointly decoding the future beam indices. Backbone reuse increases computational depth at fixed parameter count. In the standard 10-step DeepMIMO setting, R2WM achieves an average normalized beamforming gain of 0.9597, compared with 0.8629 for the prompt-as-prefix large language model, with 99.1\% fewer parameters and 59.1\% fewer counted floating-point operations. Transfer experiments show that R2WM retains normalized gain under the tested carrier-frequency and antenna-count changes, while its beam-index-only variant retains gain under the tested base-station change. Component ablations show higher gains with historical AoD conditioning and causal attention. Recursion analysis reveals a gain-computation trade-off across separately trained configurations with the same parameter count. An additional comparison with a nonrecursive Transformer of similar parameter count shows higher gain for R2WM at greater computational cost. These results support accurate long-horizon beam prediction without a large pretrained backbone.}}
\end{abstract}


\section{Introduction}
Millimeter-wave (mmWave) systems are being explored for vehicular communication and sensing \cite{LoRA-FMCW}. Large antenna arrays and highly directional beamforming compensate for severe high-frequency propagation loss. However, mobility, blockage, and changing link geometry can disrupt beam alignment, requiring frequent beam training with substantial control overhead and latency \cite{MaBeamMgmt2023,Alkhateeb2018}. \trackchange{Long-horizon beam prediction infers future beam indices from historical observations, aiming to extend beam-training intervals while supporting beam alignment.}

\leavevmode\label{rev:literature}\rev{\trackchange{Task-specific deep learning models provide compact solutions for beam prediction \cite{Guojiajia-beam-survey}. Early learning-based beam-selection methods used uplink pilots received at coordinating base stations to predict achievable rates for candidate beams and select beamforming vectors \cite{Alkhateeb2018}. Cascaded long short-term memory (LSTM) networks predict beams at discrete instants using prior low-frequency information \cite{Cascade_LSTM}. To reduce time-quantization errors, an ordinary differential equation (ODE)-LSTM predictor models continuous-time beam dynamics from received beam-training signals and predicts beams at arbitrary time instants \cite{ODE_LSTM}. These recurrent models have compact architectures, but evaluations in \cite{LiangLe} indicate limited beam-selection performance and robustness under mismatched settings.}}

\IEEEpubidadjcol
\leavevmode\label{rev:pap_lit}\rev{\trackchange{Pretrained large models have been explored across wireless tasks for their representation capacity and potential for generalization \cite{Guojiajia}. Applications include semantic communications \cite{Token-Communication}, channel acquisition \cite{CSI-GPT}, channel prediction \cite{LLM4CP2024}, and multi-tasking \cite{LLM4WM}. For beam prediction, the prompt-as-prefix large language model (PaP-LLM) adapts a frozen generative pretrained transformer 2 (GPT-2) backbone using historical beam indices and angles of departure (AoDs) \cite{LiangLe}. Although PaP-LLM outperforms recurrent baselines in the evaluated settings, its large backbone imposes substantial model-storage requirements at inference.}}
\leavevmode\label{rev:recursive_context}\rev{\trackchange{\label{excerpt:e1:start}This storage burden motivates parameter-efficient approaches that reuse a shared network for iterative refinement. An early example is the Universal Transformer, which reuses an attention-based transformation across depth \cite{UniversalTransformer}. Later, the hierarchical reasoning model (HRM) coupled two networks recurring at different frequencies \cite{HRM}. The tiny recursive model (TRM) then simplified HRM's two-network design to one shared two-layer network, reporting higher test accuracy with fewer parameters on the evaluated benchmarks \cite{TRM}. For long-horizon beam prediction, this raises the question of whether repeatedly refining representations of the future sequence can improve prediction from a fixed observation history without enlarging the model.}}

\leavevmode\label{rev:motivation}\rev{\trackchange{We therefore propose a recursive reasoning wireless model (R2WM) for long-horizon mmWave beam prediction, \label{rev:mechanism}using a shared causal backbone to refine future-beam representations conditioned on historical observations.\label{excerpt:e1:end}}}

\begin{enumerate}
\item \leavevmode\label{rev:contrib1}\rev{\trackchange{\label{excerpt:e5:start}We formulate long-horizon mmWave beam prediction as iterative refinement of future-sequence representations under a fixed observation history. R2WM repeatedly updates coupled latent and prediction states with a shared backbone before jointly decoding the future beam indices. Parameter sharing increases computational depth at fixed parameter count.\label{excerpt:e5:end}}}
\item \leavevmode\label{rev:contrib2}\rev{\trackchange{\label{excerpt:e2:start}We condition state refinement on historical AoDs and impose causal attention on the historical-future sequence. AoD embeddings provide geometric context, while the causal mask limits each future position's attention to the observation history and current or earlier future states.\label{excerpt:e2:end}}}
\item \leavevmode\label{rev:contrib3}\rev{\trackchange{\label{excerpt:e30:start}Experiments show that R2WM achieves an average normalized beamforming gain of 0.9597 over 10 steps, compared with 0.8629 for PaP-LLM, using 99.1\% fewer parameters. Separately trained recursion configurations reveal a gain-computation trade-off at fixed parameter count. Transfer experiments show that R2WM retains normalized gain under the tested carrier-frequency and antenna-count changes, while its beam-index-only variant retains gain under the tested base-station (BS) change.\label{excerpt:e30:end}}}
\end{enumerate}

\section{Problem Formulation}
\subsection{System Model}
We consider a downlink mmWave beam prediction scenario in which a single active BS serves one mobile user. The BS is equipped with an $M$-antenna uniform linear array, while the user has a single antenna. Following the geometric channel model, the downlink channel at time step $t$ is given by
\begin{equation}
\mathbf{h}_t=\sum_{l=1}^{L_t}\sqrt{\frac{1}{\rho_{t,l}}}\alpha_{t,l}\mathbf{a}(\phi_{t,l}),
\end{equation}
\leavevmode\label{rev:dimensions}\rev{where $\mathbf{h}_t,\mathbf{a}(\phi_{t,l})\in\mathbb{C}^{M\times1}$ and $L_t$ is the number of propagation paths. For path $l$, $\rho_{t,l}>0$ is the path loss, $\alpha_{t,l}\in\mathbb{C}$ is the complex gain, $\phi_{t,l}\in\mathbb{R}$ is the AoD, and $\mathbf{a}(\phi_{t,l})$ is the BS array response vector.}

\leavevmode\label{rev:path_aod}\rev{\trackchange{For channel generation, we retain the dominant line-of-sight (LOS) propagation path at each time step and use its departure azimuth as the historical AoD $\phi_t$.}}

A discrete Fourier transform (DFT) codebook with $Q$ candidate beams is adopted at the BS. The $q$-th beamforming vector is
\begin{equation}{\ifdefined\TVTmarked\color{red}\else\color{black}\fi
\begin{aligned}
\vartheta_q&=-\pi+\frac{2\pi}{Q}\left(q+\frac{1}{2}\right),\\
[\mathbf{f}_q]_m&=\frac{1}{\sqrt{M}}e^{-\mathrm{j}m\vartheta_q},
\quad m=0,\ldots,M-1,
\end{aligned}
}\end{equation}
\leavevmode\label{rev:codebook}\rev{\trackchange{where $\mathbf{f}_q\in\mathbb{C}^{M\times1}$, $\mathrm{j}^2=-1$, $q\in\{0,1,\ldots,Q-1\}$, and $\vartheta_q$ is the normalized spatial frequency of the $q$-th codeword. The\label{excerpt:e27:start} codebook may use $Q>M$, which corresponds to an oversampled spatial-frequency grid. Oversampling increases only the density of candidate spatial directions. Each codeword remains an $M$-dimensional constant-modulus beamforming vector and is therefore implementable by the $M$-element phase-shifter array.\label{excerpt:e27:end}}}

\leavevmode\label{rev:tracking}\rev{\trackchange{The following tracking procedure generates beam-index sequences used as observations and prediction targets. The first beam of each trajectory is selected by searching the complete codebook. At each subsequent time step, the search is restricted to a five-codeword neighborhood comprising the previously selected beam and its two neighbors on either side, with indices wrapping modulo $Q$. The selected index $q_t$ maximizes the gain within this neighborhood. Both\label{excerpt:e21:start} $q_t$ and the predicted index $\hat{q}_t$ identify codewords in the same $Q$-beam codebook.\label{excerpt:e21:end}}}

\rev{\trackchange{We define the normalized beamforming gain relative to the maximum over the complete codebook as}}
\begin{equation}{\ifdefined\TVTmarked\color{red}\else\color{black}\fi
G_t=
\frac{\left|\mathbf{h}_t^{\mathrm H}\mathbf{f}_{\hat{q}_t}\right|^2}
{\displaystyle\max_{q\in\{0,1,\ldots,Q-1\}}
\left|\mathbf{h}_t^{\mathrm H}\mathbf{f}_q\right|^2},
}\end{equation}
\trackchange{where $(\cdot)^{\mathrm H}$ denotes conjugate transpose}, and $G_t\in[0,1]$ measures codeword-selection quality.

\leavevmode\label{rev:history}\rev{\trackchange{At each observed time step, the model input consists of $q_t$ and the path AoD $\phi_t$.}} Given a history length $U>1$, the past $U$ beam-AoD pairs are
\begin{equation}
\{(q_{t-U+1},\phi_{t-U+1}),\dots,(q_t,\phi_t)\}.
\end{equation}
For a prediction horizon $H\geq 1$, the objective is to predict the future selected beam indices, namely,
\begin{equation}
\{q_{t+1},q_{t+2},\dots,q_{t+H}\}.
\end{equation}

\begin{figure}[!t]
\centering
\includegraphics[width=\columnwidth]{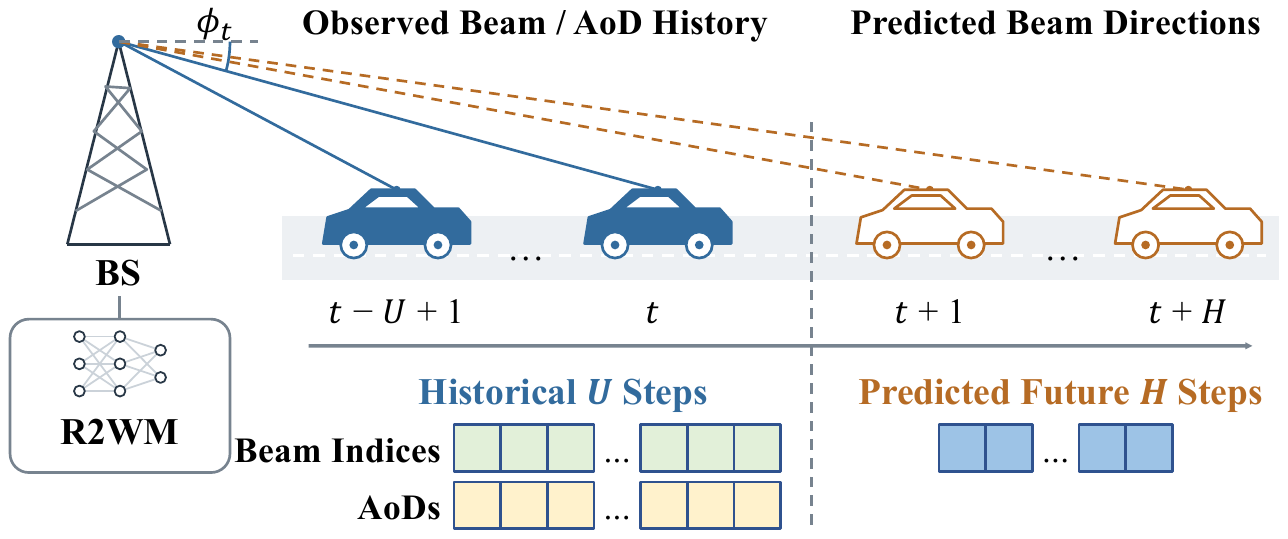}
\caption{\rev{\trackchange{Beam prediction from $U$ historical beam-index/AoD pairs. Vehicle positions show the same user at successive steps; dashed beams indicate directions selected by the $H$ predicted beam indices.}}}
\label{fig:beam_task}
\end{figure}

\rev{\trackchange{Fig.~\ref{fig:beam_task} illustrates how R2WM predicts the next $H$ beam indices from the preceding $U$ beam-index/AoD pairs.}}

\subsection{Sequence Prediction Formulation}\label{sec:sequence}
\leavevmode\label{rev:sequence}\rev{We formulate long-horizon beam prediction as masked completion of a sequence of length $U+H$. \trackchange{The first $U$ positions contain the observed beam indices, and the last $H$ positions hold placeholders for the future beam indices to be predicted.} Accordingly, the beam-index sequence is written as}
\begin{equation}
\label{ID-q}
\mathbf{q}=
[q_{t-U+1},\dots,q_t,
\underbrace{q_{\mathrm{mask}},\dots,q_{\mathrm{mask}}}_{H\ \text{positions}}]^{\mathrm T},
\end{equation}
\leavevmode\label{rev:q_mask}\rev{where $\mathbf{q}\in\mathbb{Z}^{(U+H)\times1}$ and $q_{\mathrm{mask}}$ denotes the mask token used for unknown future beam entries; $(\cdot)^{\mathrm T}$ denotes transpose. The AoD sequence is arranged in the same form,}
\begin{equation}
\label{AoD-phi}
\boldsymbol{\phi}=
[\phi_{t-U+1},\dots,\phi_t,
\underbrace{0,\dots,0}_{H\ \text{positions}}]^{\mathrm T},
\end{equation}
\leavevmode\label{rev:phi_mask}\rev{where $\boldsymbol{\phi}\in\mathbb{R}^{(U+H)\times1}$ and the last $H$ positions are set to zero because future AoDs are unavailable during inference. }

\section{Proposed R2WM Architecture}
\begin{figure*}[!t]
\centering
\includegraphics[width=0.94\textwidth]{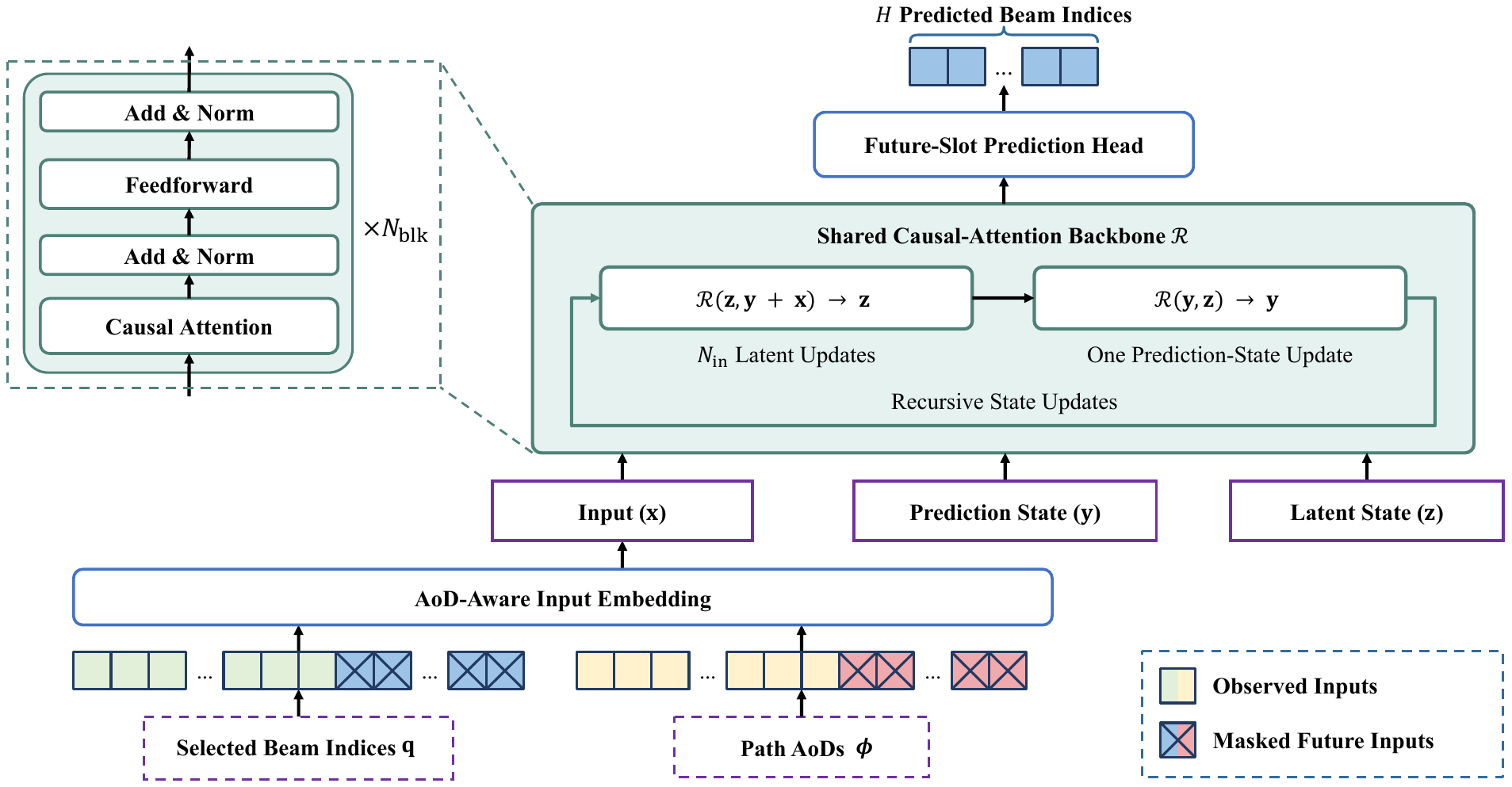}
\caption{\rev{\trackchange{R2WM architecture. Both state updates reuse the shared causal-attention backbone $\mathcal{R}$, whose internal blocks are expanded on the left. Update arrows denote assignment, with recursion indices omitted. Crossed cells denote unavailable future inputs. Only the final $H$ positions are retained for prediction after $P=16$ evaluation passes with fixed input $\mathbf{x}$.}}}
\label{model}
\end{figure*}

\leavevmode\label{rev:architecture}\rev{\trackchange{As\label{excerpt:e13:start} shown in Fig.~\ref{model}, R2WM solves the masked sequence-completion task in Section~\ref{sec:sequence} through recursive state refinement. Conditioned on fixed historical beam indices and AoDs, a shared causal backbone repeatedly updates coupled latent and prediction states before joint decoding of future beam indices.\label{excerpt:e13:end}}}

\textbf{AoD-aware input embedding.}
\leavevmode\label{rev:embedding}\rev{We embed the sequences in \eqref{ID-q}-\eqref{AoD-phi}. The beam vocabulary has size $Q+1$, where the additional token denotes the future mask. The beam-index sequence is embedded as $\mathbf{E}_q\in\mathbb{R}^{(U+H)\times d}$, where $d=512$ is the embedding dimension. \trackchange{Each scalar AoD, expressed in degrees, is divided by $90^\circ$, and the future positions are masked before projection.} \trackchange{Each normalized AoD is projected to a $d$-dimensional feature vector as}}
\begin{equation}{\ifdefined\TVTmarked\color{red}\else\color{black}\fi
\bar{\phi}_\tau
=
m_\tau\frac{\phi_\tau}{90^\circ},
\qquad
\mathbf{e}_{\phi,\tau}
=
g_{\phi}\!\left(\bar{\phi}_\tau\right),
\label{eq:aod_embedding}
}\end{equation}
\leavevmode\label{rev:embedding2}\rev{\trackchange{\trackchange{where $\tau=t-U+1,\ldots,t+H$ indexes the sequence positions, $m_\tau=1$ for the $U$ observed positions, and $m_\tau=0$ for the $H$ future positions}. The mapping $g_{\phi}:\mathbb{R}\rightarrow\mathbb{R}^{d}$ is a two-layer multilayer perceptron with a hidden dimension of 256 and sigmoid linear unit activation, followed by root mean square normalization. Future positions receive the shared embedding $g_{\phi}(0)$.} \trackchange{Stacking the projected features as rows gives} $\mathbf{E}_{\phi}\in\mathbb{R}^{(U+H)\times d}$. \trackchange{We combine the beam-index and AoD embeddings to obtain}}
\begin{equation}
\mathbf{x}=\mathbf{E}_q+\beta \mathbf{E}_{\phi},
\end{equation}
\leavevmode\label{rev:beta}\rev{where\label{excerpt:e24:start} \trackchange{$\beta=0.1$} is the injection coefficient and $\mathbf{x}\in\mathbb{R}^{(U+H)\times d}$.\label{excerpt:e24:end}}

\textbf{\trackchange{Shared causal-attention backbone.}}
\leavevmode\label{rev:backbone}\rev{The\label{excerpt:e7:start} \trackchange{shared causal-attention backbone} comprises $N_{\mathrm{blk}}=2$ stacked causal-attention blocks, each with $N_{\mathrm{head}}=8$ attention heads. \trackchange{These two blocks are evaluated on every backbone call. Causal attention acts on the entire $U+H$ sequence.}\label{excerpt:e7:end} For each head, the query, key, and value matrices satisfy $\mathbf{Q},\mathbf{K},\mathbf{V}\in\mathbb{R}^{(U+H)\times d_{\mathrm{head}}}$. The attention operation is}
\begin{equation}{\ifdefined\TVTmarked\color{red}\else\color{black}\fi
\mathrm{Attention}(\mathbf{Q},\mathbf{K},\mathbf{V})
=
\mathrm{softmax}\!\left(\frac{\mathbf{Q}\mathbf{K}^{\mathrm T}}{\sqrt{d_{\mathrm{head}}}}+\mathbf{M}\right)\mathbf{V},
}\end{equation}
\leavevmode\label{rev:mask}\rev{where $d_{\mathrm{head}}=d/N_{\mathrm{head}}$ \trackchange{is the per-head dimension. The softmax is applied row-wise.} \trackchange{The $(U+H)\times(U+H)$ additive causal mask $\mathbf{M}$ satisfies $M_{ij}=0$ for $j\leq i$ and $M_{ij}=-\infty$ otherwise, where $i$ and $j$ index sequence positions.} Each block comprises causal attention, a feedforward transformation, residual connections, and normalization. Let $\mathrm{Block}_{b}(\cdot)$ denote the $b$-th block, where $b=1,\dots,N_{\mathrm{blk}}$. For any two state tensors $\mathbf{s}_1,\mathbf{s}_2\in\mathbb{R}^{(U+H)\times d}$, the \trackchange{shared causal-attention backbone} is defined as}
\begin{equation}{\ifdefined\TVTmarked\color{red}\else\color{black}\fi
\mathcal{R}(\mathbf{s}_1,\mathbf{s}_2)
=
\mathrm{Block}_{N_{\mathrm{blk}}}\big(\cdots \mathrm{Block}_{1}(\mathbf{s}_1+\mathbf{s}_2)\cdots\big).
}\end{equation}
\leavevmode\label{rev:rope}\rev{Rotary positional embeddings are applied to the query and key representations in every causal-attention block and reused across recursive updates.}

\textbf{Recursive reasoning.}
\leavevmode\label{rev:states}\rev{\trackchange{R2WM maintains a latent state $\mathbf{z}$ for iterative refinement and a prediction state $\mathbf{y}$ for beam-index decoding, both in $\mathbb{R}^{(U+H)\times d}$. Each cycle updates $\mathbf{z}$ before $\mathbf{y}$ through the shared backbone $\mathcal{R}$:}}
\begin{equation}{\ifdefined\TVTmarked\color{red}\else\color{black}\fi
f_z(\mathbf{x}, \mathbf{y}, \mathbf{z})
=
\mathcal{R}(\mathbf{z}, \mathbf{y} + \mathbf{x}),
}\end{equation}
\begin{equation}{\ifdefined\TVTmarked\color{red}\else\color{black}\fi
f_y(\mathbf{y}, \mathbf{z})
=
\mathcal{R}(\mathbf{y}, \mathbf{z}).
}\end{equation}

\leavevmode\label{rev:recursion}\rev{\trackchange{Following\label{excerpt:e28:start} the nested TRM schedule~\cite{TRM}, let $P$ denote the number of refinement passes, $N_{\mathrm{out}}$ the number of outer prediction-state cycles per pass, and $N_{\mathrm{in}}$ the number of latent-state updates per cycle.\label{excerpt:e28:end} The pass, outer-cycle, and inner-update indices are $p=1,\ldots,P$, $r=1,\ldots,N_{\mathrm{out}}$, and $k=1,\ldots,N_{\mathrm{in}}$, respectively. Before pass $p=1$,\label{rev:state_initialization} $\mathbf{y}^{(1,0)}$ and $\mathbf{z}^{(1,1,0)}$ are broadcast from two non-trainable $d$-dimensional vectors sampled once from a variance-corrected truncated normal distribution with unit standard deviation. For $p>1$, the initial prediction and latent states equal the terminal states of pass $p-1$: $\mathbf{y}^{(p,0)}=\mathbf{y}^{(p-1,N_{\mathrm{out}})}$ and $\mathbf{z}^{(p,1,0)}=\mathbf{z}^{(p-1,N_{\mathrm{out}},N_{\mathrm{in}})}$. Within pass $p$, each cycle $r>1$ starts from $\mathbf{z}^{(p,r,0)}=\mathbf{z}^{(p,r-1,N_{\mathrm{in}})}$. The inner loop updates the latent state as}}
\begin{equation}{\ifdefined\TVTmarked\color{red}\else\color{black}\fi
\label{eq:latent_update}
\mathbf{z}^{(p,r,k)}
\leftarrow
f_z\!\left(
\mathbf{x},
\mathbf{y}^{(p,r-1)},
\mathbf{z}^{(p,r,k-1)}
\right).
}\end{equation}
The prediction state is then updated once using the final refined latent state
\begin{equation}{\ifdefined\TVTmarked\color{red}\else\color{black}\fi
\label{eq:prediction_update}
\mathbf{y}^{(p,r)}
\leftarrow
f_y\!\left(
\mathbf{y}^{(p,r-1)},
\mathbf{z}^{(p,r,N_{\mathrm{in}})}
\right).
}\end{equation}
\leavevmode\label{rev:passes}\rev{\trackchange{One\label{excerpt:e6:start} pass invokes the shared backbone $N_{\mathrm{out}}(N_{\mathrm{in}}+1)$ times. During evaluation, both states are carried through $P=16$ passes with the same fixed input $\mathbf{x}$. The default $(N_{\mathrm{out}},N_{\mathrm{in}})=(3,4)$ therefore requires $16\times3\times(4+1)=240$ backbone calls.\label{excerpt:e6:end}}}

\textbf{Future-slot prediction head.}
\trackchange{A linear output head maps the final prediction state of pass $p$ to beam-vocabulary logits}
\begin{equation}{\ifdefined\TVTmarked\color{red}\else\color{black}\fi
\boldsymbol{\ell}^{(p)}
=
\mathbf{y}^{(p,N_{\mathrm{out}})}
\mathbf{W}_{\mathrm{out}}^{\mathrm T}
+\mathbf{b}_{\mathrm{out}},
}\end{equation}
\leavevmode\label{rev:head}\rev{\trackchange{where $\mathbf{W}_{\mathrm{out}}\in\mathbb{R}^{Q\times d}$ and $\mathbf{b}_{\mathrm{out}}\in\mathbb{R}^{Q}$ are the output projection parameters. The bias is added to each row, yielding $\boldsymbol{\ell}^{(p)}\in\mathbb{R}^{(U+H)\times Q}$. Only the final $H$ rows are retained for prediction. During\label{excerpt:e18:start} training, the beam-prediction cross-entropy loss is averaged over these $H$ future positions.\label{excerpt:e18:end} At\label{excerpt:e14:start} inference, the final-pass logits $\boldsymbol{\ell}^{(P)}$ jointly yield $\hat{q}_{t+1},\ldots,\hat{q}_{t+H}$. The predicted index $\hat{q}_{t+h}$, $h=1,\ldots,H$, determines the beam used in the $h$-th subsequent transmission interval. The input window remains fixed throughout the horizon, without new observations or autoregressive beam-index feedback.\label{excerpt:e14:end}}}

\section{Simulation Results}

\subsection{System Setup}\label{sec:setup}

\leavevmode\label{rev:setup}\rev{We evaluate the proposed method on the ray-tracing-based DeepMIMO O1 scenarios \cite{DeepMIMO} in an outdoor street-canyon downlink setting. \trackchange{The standard setting uses the 28-GHz O1\_28 scenario and BS1. Carrier-frequency transfer uses the 60-GHz scenario O1\_60, and the BS-change evaluation uses BS2.} \trackchange{The BS employs a 32- or 64-element uniform linear array with half-wavelength element spacing at the corresponding carrier frequency.} \trackchange{In\label{excerpt:e26:start} the standard setting, $M=Q=64$. For antenna-mismatch evaluation, models trained in this setting are tested with $M=32$ and $Q=64$, yielding a twofold oversampled codebook.\label{excerpt:e26:end}}}

\leavevmode\label{rev:data}\rev{\trackchange{User mobility includes nominal speeds of 5, 10, 15, and 20 m/s, with instantaneous speed variations within $\pm 20\%$ of the corresponding nominal speed.} \trackchange{Each trajectory contains 100 time steps sampled at 16-ms intervals. Each sample uses $U=40$ historical steps to predict the next $H=10$ steps, spanning 160 ms.} \trackchange{The trajectories are partitioned into training and test sets before constructing windows with stride 25.} The four speed settings are mixed in both sets, yielding 82,944 training samples and 9,216 test samples.}

\begin{table}[!t]\ifdefined\TVTmarked\color{red}\else\color{black}\fi
\caption{\rev{\trackchange{Main R2WM training settings. EMA denotes exponential moving average.}}}
\label{tab:training_settings}
\centering
\small
\renewcommand{\arraystretch}{1.1}
\begin{tabular*}{\columnwidth}{@{\extracolsep{\fill}}ll@{}}
\hline
Setting & Value \\
\hline
Optimizer & AdamW \\
Learning rate & $1\times10^{-4}$ \\
Learning-rate schedule & Warmup + cosine decay \\
Batch size & 768 \\
Epochs & 200 \\
Regularization & Weight decay 0.15; dropout 0.3 \\
Evaluation passes & $P=16$ \\
Checkpoint & Final EMA at step 21,600 \\
\hline
\end{tabular*}
\end{table}

\leavevmode\label{rev:training}\rev{\trackchange{R2WM\label{excerpt:e8:start} is implemented in PyTorch and trained on a single NVIDIA RTX 4090 graphics processing unit, with the main training settings listed in Table~\ref{tab:training_settings}.\label{excerpt:e8:end} The\label{excerpt:e17:start} learning rate warms up over 4,000 optimizer steps to $1\times10^{-4}$, then follows cosine decay to $2\times10^{-5}$. For R2WM, all reported evaluations use the final exponential-moving-average (EMA) checkpoint at step 21,600.\label{excerpt:e17:end} During\label{excerpt:e25:start} preliminary development, $\beta=0.1$ was selected using training accuracy and fixed for configurations with AoD embedding, without using validation or test metrics.\label{excerpt:e25:end}}}

\leavevmode\label{rev:pap}\rev{\trackchange{We compare R2WM with three representative beam-prediction baselines: PaP-LLM~\cite{LiangLe}, CascadedLSTM~\cite{Cascade_LSTM}, and ODE-LSTM~\cite{ODE_LSTM}. Section~\ref{sec:ablation} examines the effects of AoD embedding and causal attention, analyzes different recursion configurations, and provides a separate comparison with a nonrecursive Transformer of similar parameter count.}}

\rev{\trackchange{We\label{excerpt:e19:start} implement PaP-LLM using its released codebase~\cite{LiangLe}, retaining and freezing the first 32 blocks (indices 0-31) of the 36-layer pretrained GPT-2-large backbone. The PaP pipeline processes historical beam indices and AoDs. Non-backbone modules are trained with mean-squared error for direct $H$-step prediction. After 50 training epochs, the lowest-training-loss checkpoint is used for both standard and transfer evaluations.\label{excerpt:e19:end}}}

\leavevmode\label{rev:protocol}\rev{\trackchange{The\label{excerpt:e20:start} main comparisons share the trajectory-level training/test split, window locations, $U=40$, $H=10$, future beam-index targets, and normalized beamforming-gain metric. R2WM and PaP-LLM use historical beam indices and AoDs; CascadedLSTM and ODE-LSTM use beam-training received signals~\cite{ODE_LSTM}. Future AoDs are unavailable to R2WM and PaP-LLM.\label{excerpt:e20:end}}}

\leavevmode\label{rev:timing}\rev{\trackchange{Parameter counts include frozen weights. We report per-sample FLOPs counted by \texttt{torch.profiler} over complete inference, including $P=16$ refinement passes for R2WM. The counts exclude unsupported operations such as \texttt{Conv1d}. For the models in Table~\ref{tab:hl_tradeoff}, batch-amortized inference time is measured on an RTX 4090 with batch size 768 and device synchronization, by dividing batch time by batch size.}}

\subsection{Long-Horizon Beam Prediction Results}

\begin{table}[!t]
\caption{\rev{\trackchange{Standard 64-antenna mixed-speed comparison. Parameters are in millions (M), including frozen weights. Counted FLOPs follow the profiling procedure in Section~\ref{sec:setup}.}}}
\label{tab:overall_compare}
\centering
\small
\renewcommand{\arraystretch}{1.15}
\setlength{\tabcolsep}{3.5pt}
\begin{tabular*}{\columnwidth}{@{\extracolsep{\fill}}lccc}
\hline
Method & Params. (M) & \makecell{FLOPs\\(GFLOPs/sample)} & Avg. gain \\
\hline
CascadedLSTM & 0.791 & 0.092 & 0.8313 \\
ODE-LSTM     & 0.922 & 0.449 & 0.8490 \\
PaP-LLM      & 749.632 & 528.352 & 0.8629 \\
R2WM         & 7.016 & 216.204 & \textbf{0.9597} \\
\hline
\end{tabular*}
\end{table}

\begin{figure}[!t]
\centering
\includegraphics[width=0.85\columnwidth]{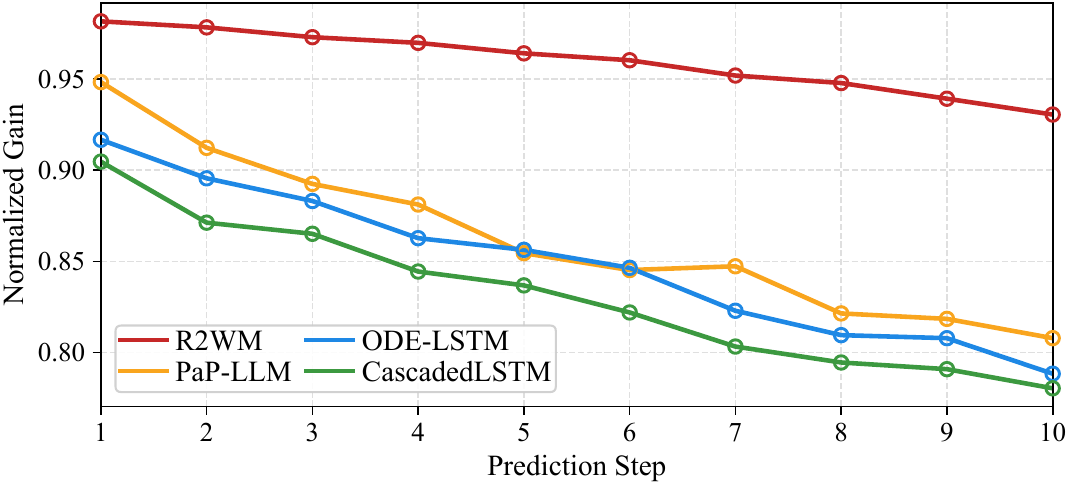}
\caption{\rev{Comparison of normalized beamforming gain over a 10-step prediction horizon under a 64-antenna configuration with mixed user speeds.}}
\label{fig:results}
\end{figure}

\leavevmode\label{rev:results}\rev{\trackchange{As\label{excerpt:e9:start} shown in Table~\ref{tab:overall_compare}, R2WM achieves a 10-step average normalized gain of 0.9597 under the standard 64-antenna mixed-speed setting, compared with 0.8629 for PaP-LLM.\label{excerpt:e9:end} At the final prediction step, R2WM achieves a normalized gain of 0.9306, compared with 0.8079 for PaP-LLM, 0.7884 for ODE-LSTM, and 0.7804 for CascadedLSTM (Fig.~\ref{fig:results}). R2WM\label{excerpt:e11:start} uses 99.1\% fewer parameters and 59.1\% fewer counted FLOPs than PaP-LLM. The recurrent baselines require fewer parameters and FLOPs than R2WM but achieve lower normalized gains.\label{excerpt:e11:end}}}

\begin{figure}[!t]
\centering
\includegraphics[width=0.85\columnwidth]{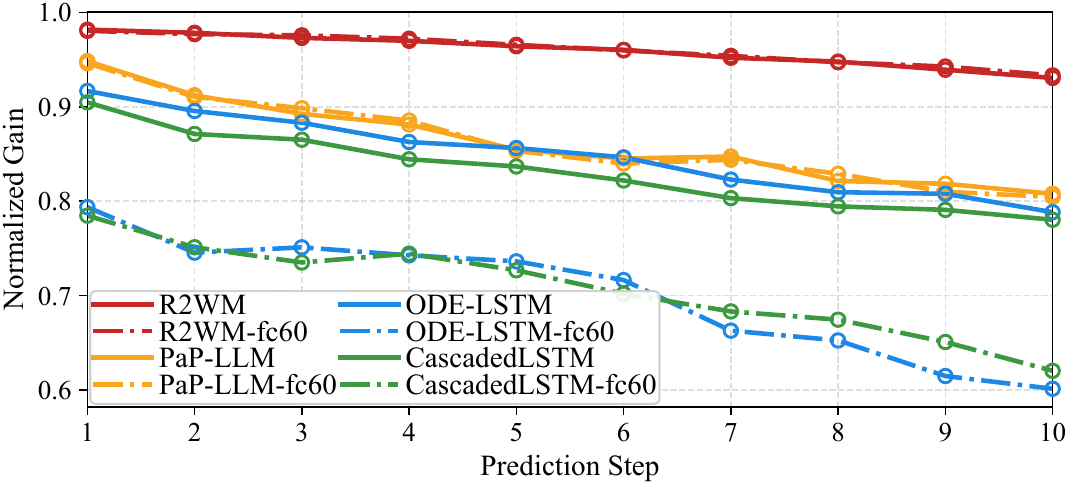}
\caption{\rev{\trackchange{Transfer from 28 to 60 GHz without retraining, using complete R2WM with $M=Q=64$ and half-wavelength spacing at each frequency.}}}
\label{fig:freq_generalization}
\end{figure}

\leavevmode\label{rev:frequency}\rev{\trackchange{Fig.~\ref{fig:freq_generalization} evaluates direct transfer from 28 GHz to 60 GHz without retraining. R2WM attains average normalized gains of 0.9597 at 28 GHz and 0.9608 after transfer, while PaP-LLM attains 0.8629 and 0.8621, respectively. Both\label{excerpt:e15:start} scenarios use half-wavelength element spacing at their respective carrier frequencies and the same 64-codeword DFT grid in (2) for a 64-element array. In\label{excerpt:e10:start} the single-path setting, normalization removes the effect of overall channel strength, so the gain reflects how well the selected beam aligns with the path direction. At a given angle, half-wavelength spacing and the common DFT grid preserve the beam-index mapping and normalized codeword gains across frequencies. The nearly unchanged normalized gains indicate that the trained model retains its beam-selection performance across these frequency settings with the same normalized array geometry and codebook.\label{excerpt:e10:end}\label{excerpt:e15:end}}}

\begin{figure}[!t]
\centering
\includegraphics[width=0.85\columnwidth]{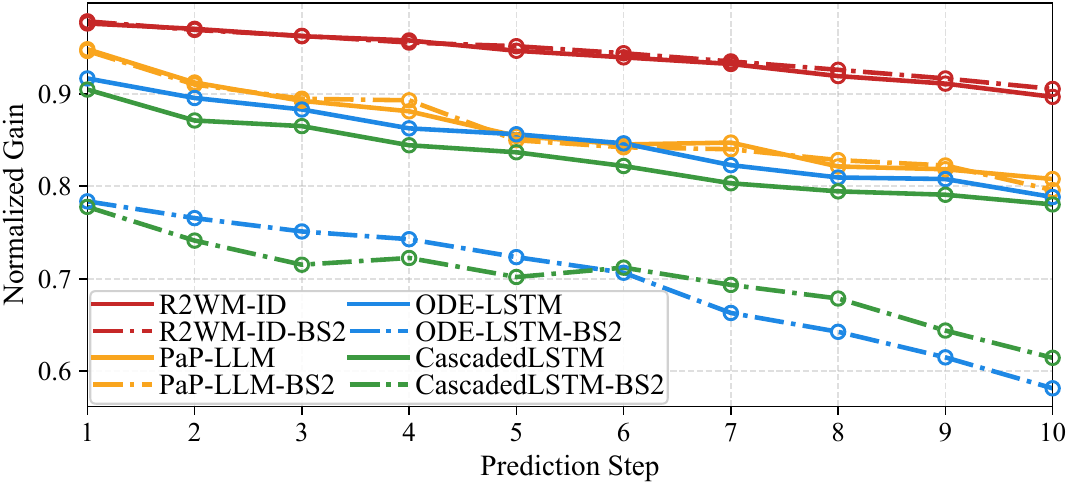}
\caption{\rev{\trackchange{Robustness to a BS change from BS1 to BS2 without retraining. R2WM-ID uses beam-index history only.}}}
\label{fig:bs_generalization}
\end{figure}

\leavevmode\label{rev:bs}\rev{\trackchange{We\label{excerpt:e22:start} next evaluate robustness to a BS change from BS1 to BS2 without retraining using R2WM-ID, the beam-index-only variant without AoD embedding (Fig.~\ref{fig:bs_generalization}).\label{excerpt:e22:end} Its 10-step average gain is 0.9413 at BS1 and 0.9445 after transfer to BS2, with corresponding step-10 gains of 0.8968 and 0.9054. These similar gains indicate that R2WM-ID retains normalized gain under the tested BS change.}}

\begin{figure}[!t]
\centering
\includegraphics[width=0.85\columnwidth]{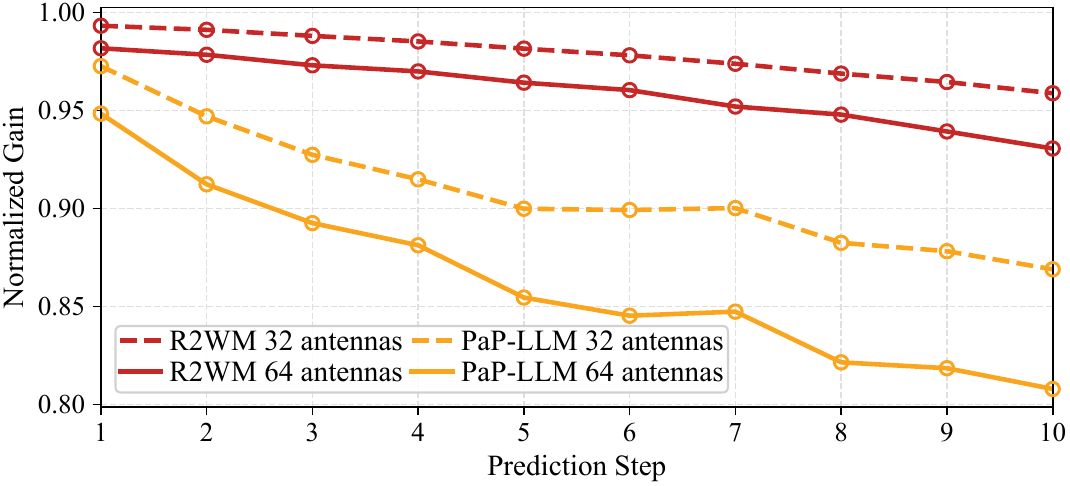}
\caption{\rev{\trackchange{Transfer of complete R2WM from $M=64$ to $M=32$ with $Q=64$ fixed. Gain is normalized separately for each physical array.}}}
\label{fig:antenna_generalization}
\end{figure}

\leavevmode\label{rev:antenna}\rev{\trackchange{Fig.~\ref{fig:antenna_generalization} compares matched evaluation at $M=64,Q=64$ with antenna-mismatch evaluation at $M=32,Q=64$. R2WM achieves average normalized gains of 0.9597 in the matched setting and 0.9783 in the antenna-mismatch setting. The corresponding gains for PaP-LLM are 0.8629 and 0.9090. The\label{excerpt:e3:start} same 64-codeword vocabulary allows the trained predictor to retain its output classes when the antenna count changes. The wider beams of the smaller array may reduce the normalized-gain penalty caused by a given pointing error. Normalization uses the maximum codebook gain for each array. The reported values therefore measure relative beam-selection quality rather than absolute array gain.\label{excerpt:e3:end}}}

\subsection{\trackchange{Ablation and Recursion Analysis}}\label{sec:ablation}

\begin{figure}[!t]
\centering
\includegraphics[width=0.85\columnwidth]{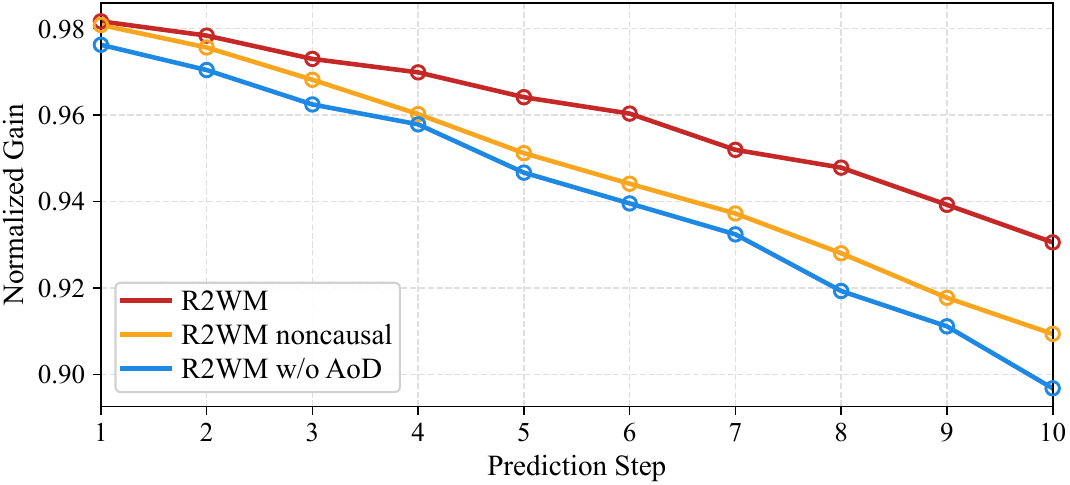}
\caption{\rev{\trackchange{\trackchange{Effects of removing AoD embedding and replacing the full-sequence causal mask with noncausal attention.} Future inputs remain masked in both attention variants.}}}
\label{fig:ablation_components}
\end{figure}

\leavevmode\label{rev:components}\rev{\trackchange{Fig.~\ref{fig:ablation_components} evaluates the AoD embedding and temporal attention constraint. Removing AoD embedding reduces the average gain from 0.9597 to 0.9413 and the step-10 gain from 0.9306 to 0.8968, supporting historical AoD conditioning. Separately, replacing causal attention with noncausal attention reduces the full model's average and step-10 gains to 0.9473 and 0.9094. Both\label{excerpt:e4:start} attention variants receive mask tokens for future beam indices and zeros for future AoDs. Causal attention limits each future position's attention to the observed history and current or earlier future states. The gain difference supports this temporal constraint.\label{excerpt:e4:end}}}

\begin{figure}[!t]
\centering
\includegraphics[width=0.85\columnwidth]{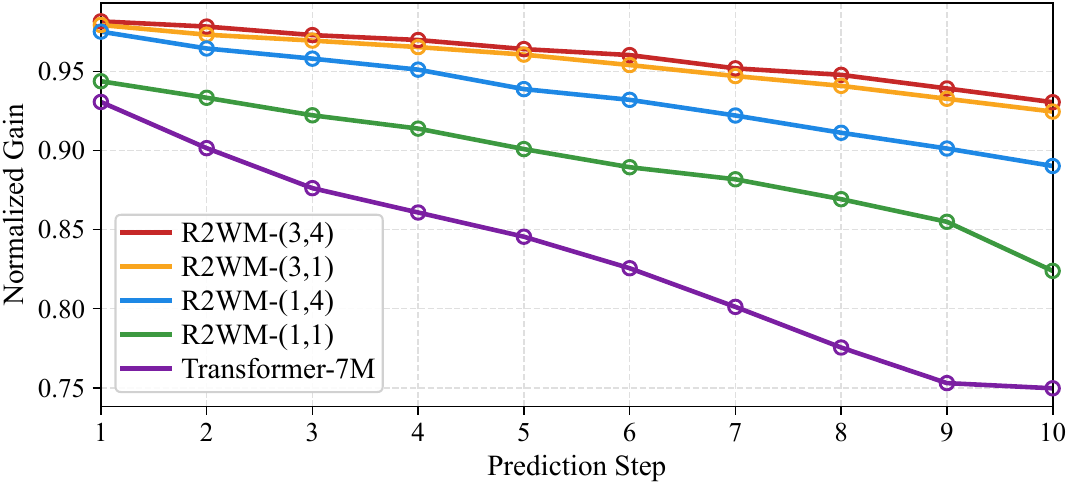}
\caption{\rev{\trackchange{R2WM recursion configurations and a nonrecursive Transformer with approximately 7M parameters. The R2WM configurations use $P=16$ evaluation passes.}}}
\label{fig:ablation_recursion}
\end{figure}

\leavevmode\label{rev:recursion_gain}\rev{\trackchange{\label{excerpt:e29:start}Fig.~\ref{fig:ablation_recursion} compares the default R2WM-$(3,4)$, three reduced-recursion variants, $(3,1)$, $(1,4)$, and $(1,1)$, and a nonrecursive Transformer reference. Each R2WM configuration is trained separately with 7.016M parameters. Reducing the recursion configuration from $(3,4)$ to $(1,1)$ lowers average normalized gain from 0.9597 to 0.8933 and step-10 gain from 0.9306 to 0.8239. The final-step gain reduction exceeds the horizon-average reduction.\label{excerpt:e29:end}}}

\leavevmode\label{rev:tradeoff}\rev{\trackchange{Table~\ref{tab:hl_tradeoff} reports inference costs for the four R2WM configurations and Transformer-7M; only R2WM uses $P=16$ evaluation passes. \trackchange{Compared with $(3,4)$, $(3,1)$ reduces batch-amortized time by 58.1\% and FLOPs by 59.9\%, with an average gain reduction of 0.0049.} \label{excerpt:e16:start}\trackchange{Thus, $(3,1)$ reduces computation, whereas $(3,4)$ achieves the highest normalized gain among the tested configurations. These timings are batch-amortized; single-user and end-to-end BS-side latency remain to be evaluated.}\label{excerpt:e16:end}}}

\leavevmode\label{rev:transformer_setup}\rev{\trackchange{\label{excerpt:e31:start}Transformer-7M has 7.081M parameters, eight bidirectional encoder layers, one causal decoder layer, model dimension 256, eight attention heads, and feedforward dimension 960. Its decoder receives the last ten observed beam indices and ten zero placeholders. The model is trained for 50 epochs with Adam, mean-squared-error loss, and batch size 32. A one-cycle learning-rate schedule peaks at $10^{-3}$; gain evaluation uses the lowest-training-loss checkpoint.\label{excerpt:e31:end}}}

\leavevmode\label{rev:transformer_comparison}\rev{\trackchange{\label{excerpt:e32:start}R2WM and Transformer-7M differ in architecture and training objective; their comparison evaluates complete predictors of similar parameter count rather than isolating recursion. R2WM-$(3,4)$ achieves average and step-10 normalized gains of 0.9597 and 0.9306, compared with 0.8319 and 0.7496 for Transformer-7M. Transformer-7M uses fewer counted FLOPs and shorter batch-amortized inference time (Table~\ref{tab:hl_tradeoff}).\label{excerpt:e32:end}}}

\begin{table}[!t]
\caption{\rev{\trackchange{Normalized beamforming gain and inference cost. R2WM settings are $(N_{\mathrm{out}},N_{\mathrm{in}})$; $P=16$. Batch-amortized time is in ms/sample and counted FLOPs in GFLOPs/sample.}}}
\label{tab:hl_tradeoff}
\centering
\small
\renewcommand{\arraystretch}{1.15}
\setlength{\tabcolsep}{3.5pt}
\begin{tabular*}{\columnwidth}{@{\extracolsep{\fill}}lcccc}
\hline
\trackchange{Method} & \trackchange{Params. (M)} & \trackchange{Time/sample} & FLOPs & Avg. gain \\
\hline
\trackchange{R2WM-(3,4)} & 7.016 & 2.228 & 216.204 & 0.9597 \\
\trackchange{R2WM-(3,1)} & 7.016 & 0.933 & 86.651 & 0.9548 \\
\trackchange{R2WM-(1,4)} & 7.016 & 0.789 & 72.256 & 0.9345 \\
\trackchange{R2WM-(1,1)} & \trackchange{7.016} & \trackchange{0.349} & \trackchange{29.071} & \trackchange{0.8933} \\
\trackchange{Transformer-7M} & \trackchange{7.081} & \trackchange{0.017} & \trackchange{0.209} & \trackchange{0.8319} \\
\hline
\end{tabular*}
\end{table}

\section{Conclusion}
\leavevmode\label{rev:conclusion}\rev{\trackchange{\trackchange{\label{excerpt:e12:start}In this paper, we have proposed R2WM for long-horizon mmWave beam prediction by recursively refining coupled states with a shared causal backbone. In the standard 10-step setting, R2WM achieves average normalized gain of 0.9597, compared with 0.8629 for PaP-LLM, using 99.1\% fewer parameters and 59.1\% fewer counted FLOPs. \label{excerpt:e23:start}Component ablations support historical AoD conditioning and causal attention. Separately trained recursion configurations reveal a gain-computation trade-off at fixed parameter count.\label{excerpt:e23:end} R2WM retains normalized gain under the tested carrier-frequency and antenna-count changes, and R2WM-ID retains normalized gain under the tested BS change. Future work will investigate end-to-end inference latency on BS hardware and extensions to multi-user and multi-BS scenarios.\label{excerpt:e12:end}}}}

\ifdefined\TVTmarked
\let\trackoriginalbibitem\bibitem
\renewcommand{\bibitem}[1]{%
  \color{black}%
  \ifnum\pdfstrcmp{#1}{UniversalTransformer}=0\color{red}\fi
  \ifnum\pdfstrcmp{#1}{HRM}=0\color{red}\fi
  \trackoriginalbibitem{#1}%
}
\fi
\bibliographystyle{IEEEtran}
\bibliography{ref}

\vfill

\end{document}